\documentclass[a4paper,11pt]{article}
\pdfoutput=1 % if your are submitting a pdflatex (i.e. if you have
\usepackage{jinstpub} % for details on the use of the package, please
\usepackage{lineno}
\usepackage{caption}
\usepackage{subcaption}
\title{Comparison of the measured atmospheric muon rate with Monte Carlo simulations and sensitivity study for detection of prompt atmospheric muons with KM3NeT}

\author[a,1]{P. Kalaczyński,\note{Corresponding author.}}

\affiliation[a]{National Centre for Nuclear Research,\\Pasteura 7, Warsaw, Poland}

\emailAdd{piotr.kalaczynski@ncbj.gov.pl}

\abstract{The KM3NeT Collaboration has successfully deployed the first detection units of the next generation undersea neutrino telescopes in the Mediterranean Sea at the two sites in Italy and in France. A sample of the data collected between December 2016 and January 2020 has been used to measure the atmospheric muon rate at two different depths under the sea level: at 3.5 km with KM3NeT-ARCA and at 2.5 km with KM3NeT-ORCA. Atmospheric muons represent the dominant signal in a neutrino telescope and can be used to test the reliability of the Monte Carlo simulation chain and to study the physics of extensive air showers caused by highly-energetic primary nuclei impinging the Earth’s atmosphere. At energies above PeV the contribution from prompt muons, created right after the first interaction in the shower, is expected to become dominant, however, its existence has not yet been experimentally confirmed. In this work, data collected with the first detection units of KM3NeT are compared to Monte Carlo simulations based on MUPAGE and CORSIKA codes. The main features of the simulation and reconstruction chains are presented. Additionally, the first results of the simulated signal from the prompt muon component for KM3NeT-ARCA and KM3NeT-ORCA obtained with CORSIKA  are discussed.}


\collaboration[c]{on behalf of the KM3NeT Collaboration}

\proceeding{VLVnT2021\\
  18-21 May 2021\\
  Valencia}

\begin{document}
\maketitle
\flushbottom

\section{Introduction}

KM3NeT research infrastructure consists of two neutrino detectors under construction at the bottom of the Mediterranean Sea: ARCA (Astroparticle Research with Cosmics in the Abyss) located off-shore Portopalo di Capo Passero, Sicily, Italy, at a depth of 3500\,m and ORCA (Oscillation Research with Cosmics in the Abyss) off-shore Toulon, France, at a depth of 2450\,m. 
%More detailed description of detector design and physics goals can be found in \cite{LoI}.
ARCA is designed to observe TeV-PeV cosmic neutrinos and identify their astrophysical sources. ORCA focuses on the study of GeV atmospheric neutrino oscillations, in order to determine the neutrino mass hierarchy (NMH) \cite{LoI}. This translates roughly to similar muon energies and to primary cosmic ray (CR) energies in PeV-EeV range for ARCA and TeV for ORCA (although the detectors are in fact sensitive to a wider energy range). The angular acceptances of full ARCA and ORCA detectors are best for horizontal muons and slightly decrease towards the vertical direction.

Both detectors consist of vertically aligned detection units (DUs), with 18 digital optical modules (DOMs) on each DU \cite{LoI}. Every DOM contains 31 3-inch photomultiplier tubes (PMTs), calibration and positioning instruments and readout electronics boards. ARCA and ORCA have different horizontal (90\,m and 20\,m respectively) and vertical (36\,m and 9\,m respectively) spacing between the DOMs, since they are optimised for different energy ranges. In their final configuration, there will be 115 DUs at ORCA and 2x115 DUs (in two blocks) at ARCA site. 

The KM3NeT Monte Carlo (MC) simulations of CR-muons are based on two event generators: MUPAGE \cite{MUPAGE} and CORSIKA \cite{CORSIKA}. 
% MUPAGE is a fast simulation program, based on parametric formulas. It generates muon bundles induced by CRs impinging the Earth’s atmosphere. The distributions of muons are sampled directly on the surface of the active volume of the detector at different undersea depths and zenith angles. CORSIKA simulates the interactions of the primary CRs in the atmosphere and follows the shower development to the sea level. CORSIKA allows to choose between different hadronic interaction and primary CR composition models. 
Propagation of CORSIKA muons from the sea level to the active volume of the detector is done by the gSeaGen code \cite{gSeaGen} with a chosen 3-dimensional muon propagator, PROPOSAL \cite{PROPOSAL}. Next, Cherenkov photons emitted by the water molecules due to the passage of the muons and their detection by the DOMs are simulated with a custom application, called JSirene \cite{JSirene}. Afterwards, the environmental optical background (photons due to bioluminescence and to \(^{40}\mathsf{K}\) decays) is added and front-end electronics response is simulated. The MUPAGE simulation is performed in a run-by-run mode, i.e. a simulated run is produced for each data run. Conversely, average settings over all runs are used for CORSIKA MC. Trigger algorithms, as used for real data, are applied to identify possible interesting events in the simulated sample \cite{AtmMuRate}. Finally, the direction and energy of the selected events are reconstructed based on the hit information from the PMTs with an algorithm "JGandalf" used for the real data stream as well \cite{reco}. For muons above 10 TeV, the angular resolution is 0.2\(^{\circ}\) and the energy resolution is 0.28 in units of \(\log_{10}(E_{\mu})\) \cite{LoI}. At this point the simulated MC events are compared to the calibrated experimental data.

Atmospheric muons are the most abundant signal in a neutrino telescope and pose a major background for physics analyses with neutrinos. Moreover, muon data is invaluable for testing the detector performance and validation of the Monte Carlo simulations. In this work, the first data from ARCA2 (ARCA with 2 DUs) and ORCA4 (ORCA with 4 DUs) is compared to MC simulations.

Muons convey information about extensive air showers (EAS) and the CR primaries causing them. For muons with PeV and larger energies, the dominant muon yield is expected to come from the decay of short-lived hadrons. Such muons are called prompt muons, opposed to the conventional muons produced mostly in decays of charged pions and kaons. The existence of the prompt muon flux has not been yet experimentally confirmed, however, there is some evidence from IceCube (strongly dependent on the choice of the primary CR flux model) \cite{ICECUBE_PROMPT}. The first results of a simulation study of the prompt muon flux component at KM3NeT detectors are shown in Section \ref{prompt-ana}.

What KM3NeT detectors often observe are not single, but multiple muons. Hence, in the following we will use the concept of muon bundles, which are multi-muon events. A single muon event is a bundle with multiplicity equal to one.

\section{Data vs MC comparisons}
\label{sec:data-vs-MC}

% zenith plots
\begin{figure}
     \centering
     \begin{subfigure}{0.49\textwidth}
         \centering
         \includegraphics[width=\textwidth]{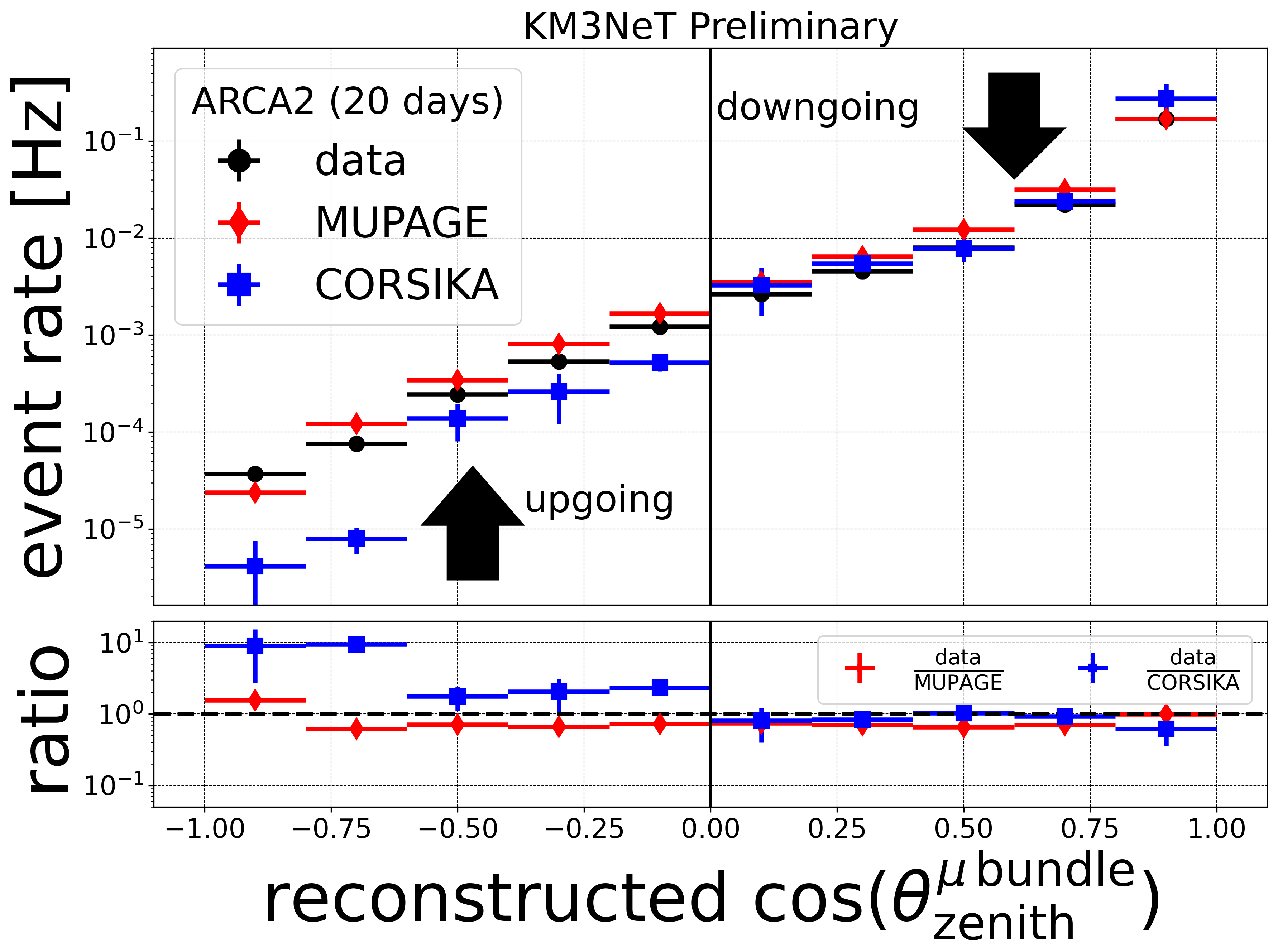}
         \caption{Distribution for ARCA2.}
         \label{fig:zenith-1}
     \end{subfigure}
     \hfill
     \begin{subfigure}{0.49\textwidth}
         \centering
         \includegraphics[width=\textwidth]{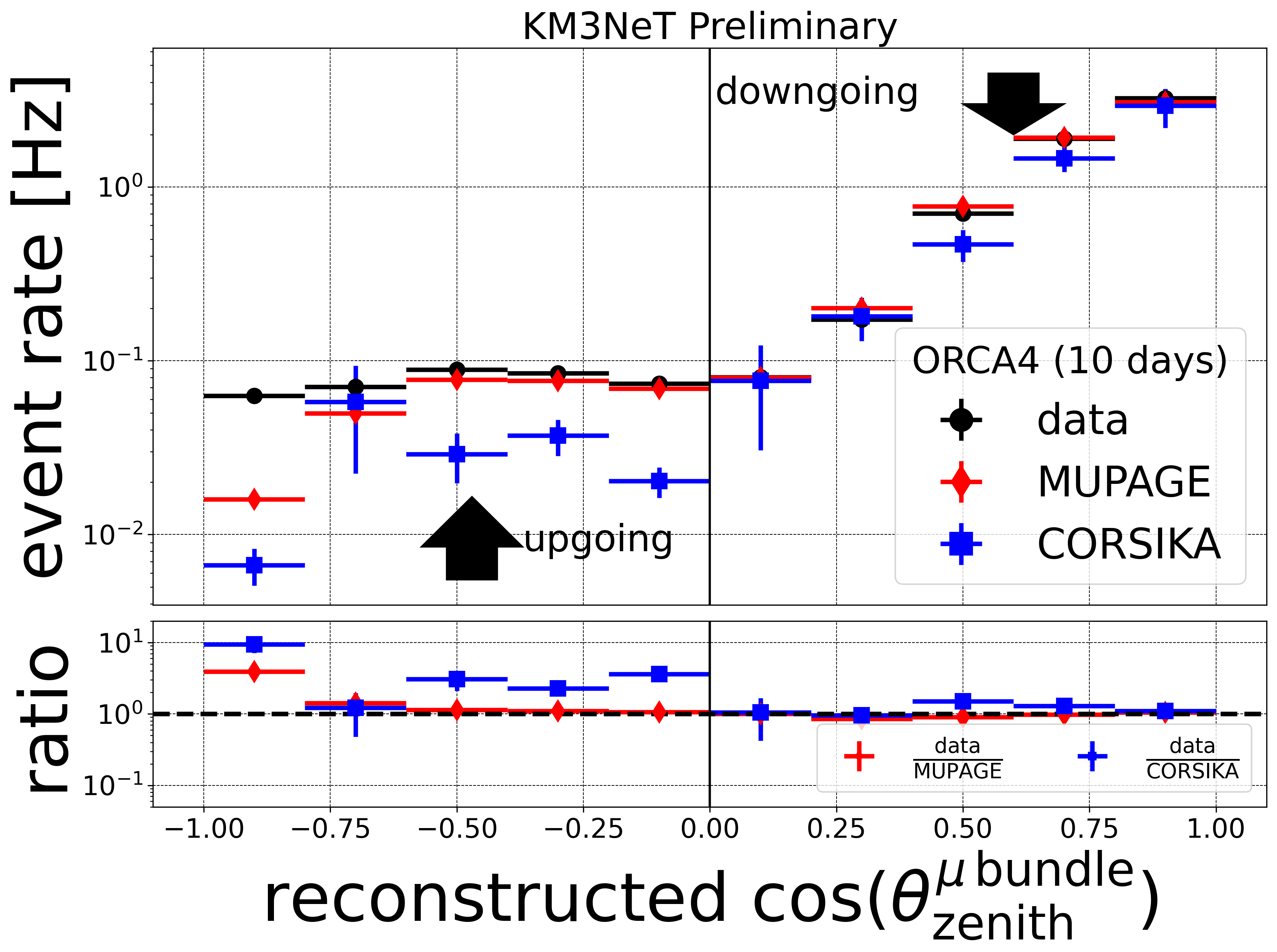}
         \caption{Distribution for ORCA4.}
     \label{fig:zenith-2}
     \end{subfigure}
        \caption{Rate of atmospheric muons as a function of the reconstructed zenith angle for data and MC simulation for the ARCA2 and ORCA4 detector. Most of the upgoing events (all MC events) are downgoing events that are badly reconstructed as upgoing. No quality cuts have been applied to remove the badly reconstructed tracks.}
        \label{fig:zeniths}
\end{figure}

% energy plots
\begin{figure}
     \centering
     \begin{subfigure}[b]{0.49\textwidth}
         \centering
         \includegraphics[width=\textwidth]{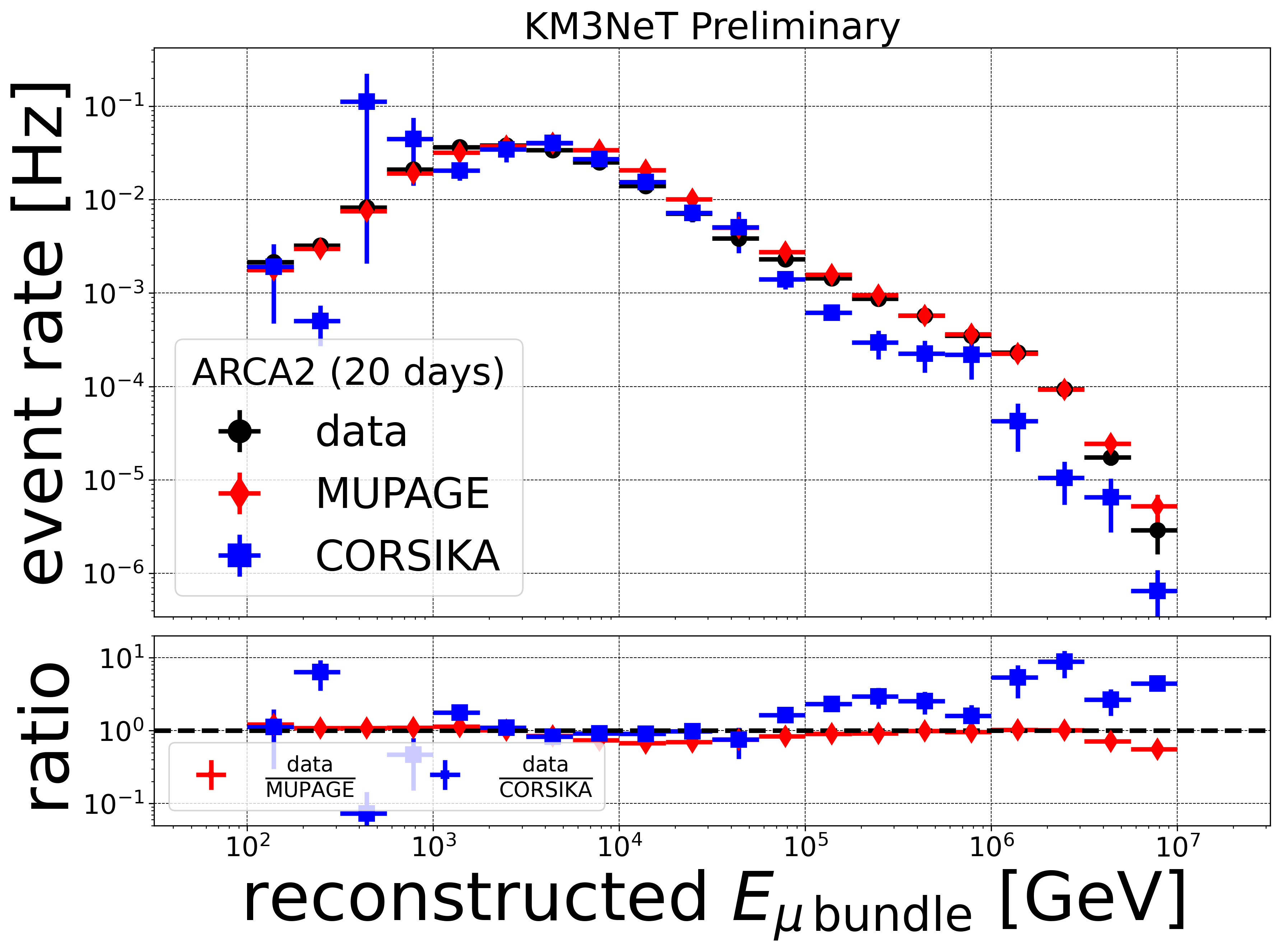}
         \caption{Distribution for ARCA2.}
         \label{fig:energy-1}
     \end{subfigure}
     \hfill
     \begin{subfigure}[b]{0.49\textwidth}
         \centering
         \includegraphics[width=\textwidth]{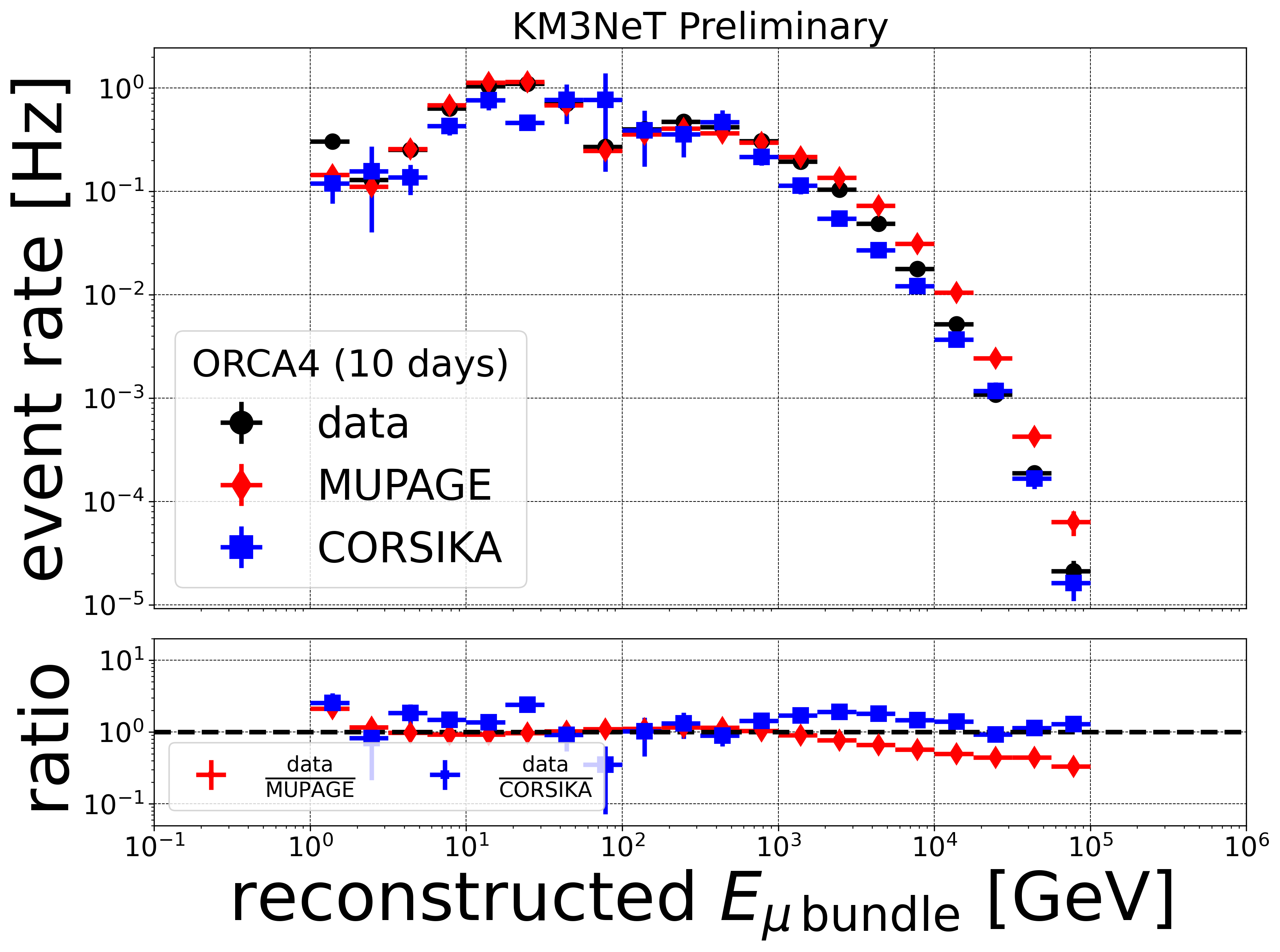}
         \caption{Distribution for ORCA4.}
     \label{fig:energy-2}
     \end{subfigure}
        \caption{Rate of atmospheric muons as a function of the reconstructed energy for data and MC simulation  for the ARCA2 and ORCA4 detector. More details on the energy reconstruction can be found in \cite{reco}.}
        \label{fig:energies}
\end{figure}

A selection of data runs (each run lasts 6 hours) collected with ARCA2 from the 23th of December 2016 to the 2nd of March 2017 and with ORCA4 from the 23th of July to 16th of December 2019 has been compared with the expectation of a MC simulation based on identical triggering and reconstruction programs.

A sample of muon bundles with energies above 10 GeV and multiplicities up to 100 muons per bundle was produced with MUPAGE. The equivalent livetime is about 20\,days for ARCA2 and 10\,days for ORCA4, which is similar to the true detector livetimes.

In total, \(2.5\cdot10^9\) showers have been simulated with CORSIKA, using SIBYLL-2.3c as the high-energy hadronic interaction model \cite{SIBYLL}. The simulated CR primaries were: \(p\), \(He\), \(C\), \(O\) and \(Fe\) nuclei with energies between 1\,TeV and 1\,EeV. The events were weighted assuming the GST3 CR composition model \cite{GST}. All events, real and simulated, were reconstructed with the same algorithm (JGandalf). 

In Figures \ref{fig:zeniths} and \ref{fig:energies} the results of the simulation with MUPAGE and CORSIKA are represented in red and blue respectively. Data are black. Only statistical errors are indicated. No systematic uncertainties are considered in the plots, however, they are expected to be larger than the statistical errors. For CORSIKA, statistical errors are evaluated as  \(\Delta x=\sqrt{{\sum}w_{i}^{2}}\), where \(w_{i}\) is the weight of the \(i\)-th MC event.

The presented results show that the MC simulations match the data where the muon flux peaks for each detector (downgoing, 1-10 TeV for ARCA and 10 GeV - 1 TeV for ORCA) and reproduce the general shape of the distributions, although there is certainly room for improvement. The muons reconstructed as upgoing in Figure \ref{fig:zeniths} are misreconstructed downgoing muons. No quality selection has been applied to remove these poorly reconstructed tracks. The difference between muons from CORSIKA and data/MUPAGE reconstructed as upgoing was identified as coming from the errors in processing in gSeaGen and will be fixed in the future. The energy distributions in Figure \ref{fig:energies} show significant deviation from the data at highest bundle energies. It is under investigation why it occurs and why only for CORSIKA in ARCA2 and only for MUPAGE in ORCA4 plot.

\section{Prompt muon analysis}\label{prompt-ana}

The muon flux is commonly divided into conventional and prompt components. The majority of conventional muons come from $\pi^{\pm}$ and $K^{\pm}$ decays, whereas the prompt muons are created in decays of heavy hadrons and light vector mesons. The potential of KM3NeT to observe the prompt muon flux is investigated in this work.

The prompt muon analysis is performed with a CORSIKA MC sample containing \(5.3\cdot10^6\) simulated showers and SIBYLL-2.3d as high-energy hadronic interaction model \cite{SIBYLL-2.3d} (version 2.3d includes corrections potentially relevant to the analysis). The same groups of CR nuclei as in Section \ref{sec:data-vs-MC} are used as primaries, however in the energy range from 0.9\,PeV up to 40\,EeV. Primaries with lower energies produce muons well below the expected signal region.

\begin{figure}
     \centering
     \begin{subfigure}[b]{0.49\textwidth}
         \centering
         \includegraphics[width=\textwidth]{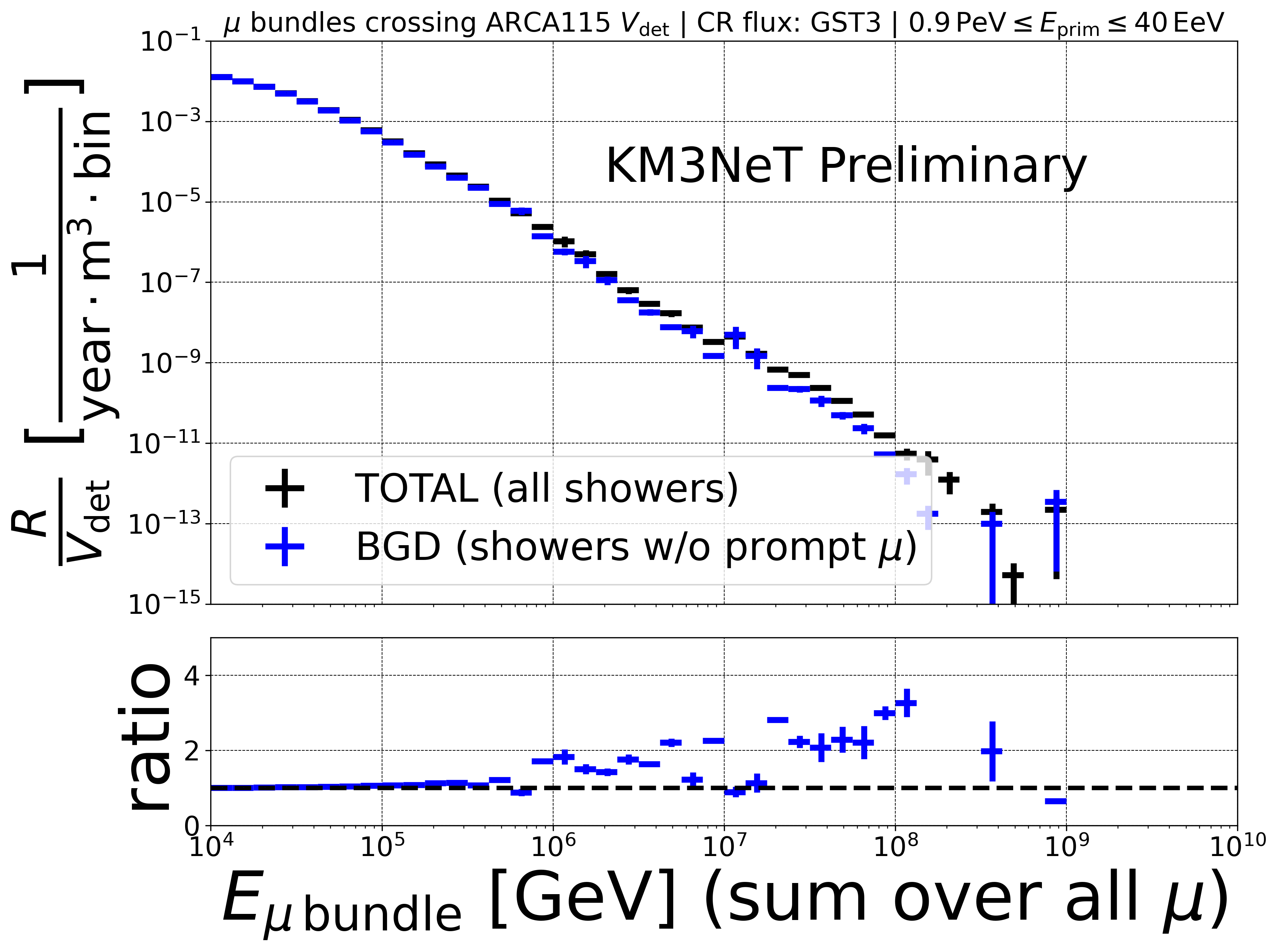}
         \caption{Distribution of muon bundle energies (sum of energies of individual muons).}
         \label{fig:prompt-1}
     \end{subfigure}
     \hfill
     \begin{subfigure}[b]{0.49\textwidth}
         \centering
         \includegraphics[width=\textwidth]{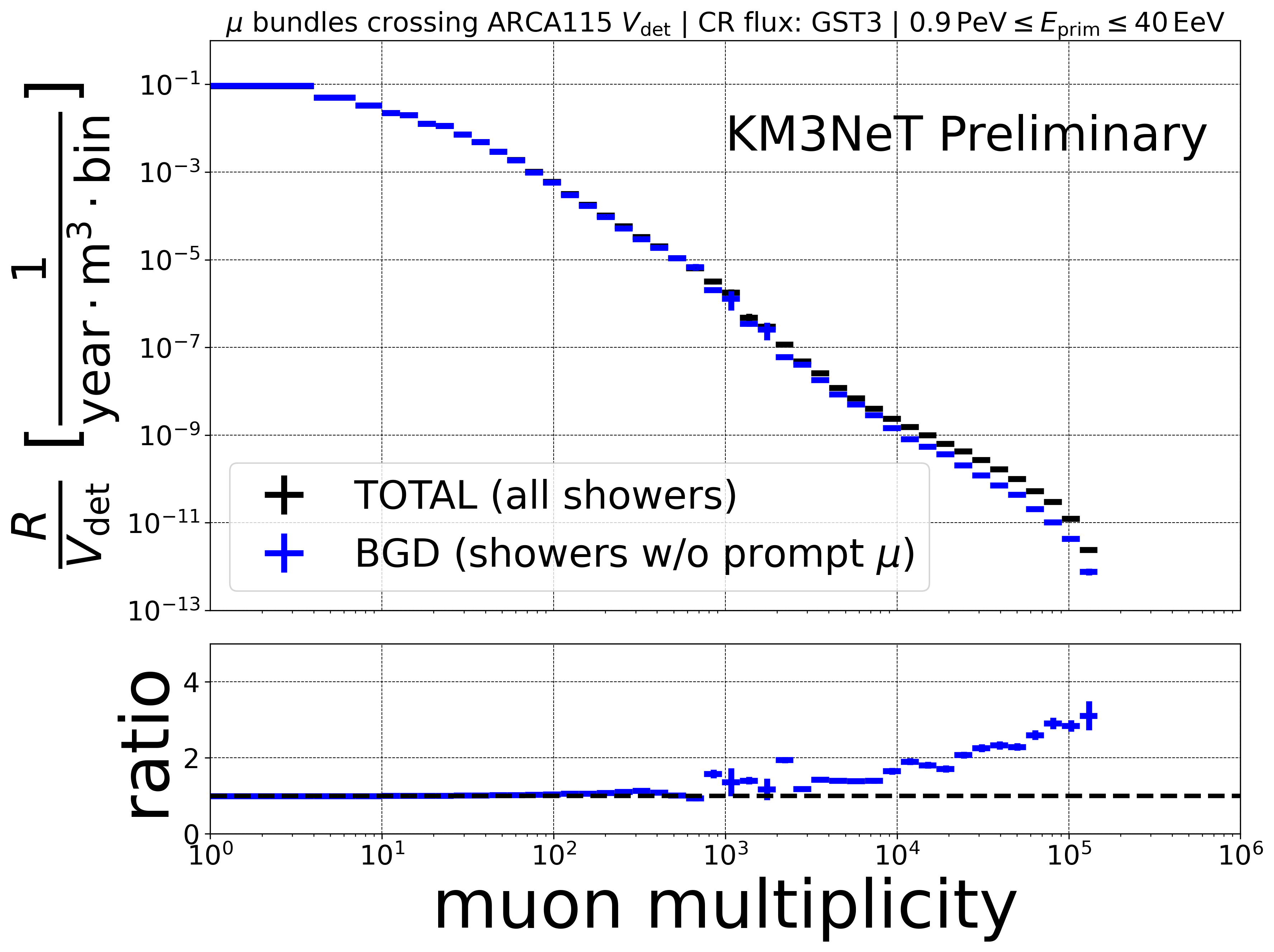}
         \caption{Distribution of muon multiplicities (number of muons in a bundle).}
     \label{fig:prompt-2}
     \end{subfigure}
        \caption{Rate of atmospheric muon bundles per detector volume as a function of the bundle energy or multiplicity for BGD and TOTAL for the ARCA115 detector.}
        \label{fig:prompt}
\end{figure}

The definition of prompt muons was introduced requiring that all muon parent particles have lifetimes shorter than $K_{\mathsf{S}}^{0}$ ($8.95\cdot10^{-11}$s) and that the muon has no more than two parent generations (mother and grandmother). The signal (SIG) for the analysis is defined as muon bundles with at least one prompt muon. The bundles with no prompt muons are considered a background (BGD). All muon bundles together are referred to as TOTAL. The distributions of muon bundle energies and multiplicities for BGD and TOTAL at the detector volume for ARCA115 (one building block of ARCA; 115 Detection Units installed) are shown in Figure \ref{fig:prompt}. There is a clear excess above BGD starting around 1\,PeV for the energy distribution and above \(10^3\) in the case of multiplicity distribution. This is a promising prediction, even though the differences between BGD and TOTAL are not large. Further investigation is required at the reconstruction level and the impact of the systematic uncertainties has to be assessed.

\section{Conclusions}

A comparison of the measured and the expected muon rate is possible using data from the first bunches of ARCA and ORCA detection units in operation. The comparisons between data and MC provide consistent results, however, there are ongoing efforts to further improve their agreement.
The analysis will be repeated with ARCA6 and ORCA6 (ARCA/ORCA with six Detection Units installed) data and with new MUPAGE and CORSIKA simulations. Several improvements are foreseen in the MC chain: optimisation of the muon propagation step, improved definition of the geometry of muon propagation, a more accurate event weight calculation and inclusion of the delta ray contribution in the energy reconstruction. A systematic uncertainty study and an increase of the statistical significance of the sample is foreseen for the CORSIKA MC production.

The first estimation of the expected signal from prompt muons is encouraging and suggests that KM3NeT may be sensitive enough to measure the prompt component. Figure \ref{fig:prompt} presents the expected result for ARCA115. 
% Similar results have been obtained also for ORCA detector, however due to space limitations, they could not be presented here. 
Further investigations at the reconstruction level are required before starting a comparison with IceCube results \cite{ICECUBE_PROMPT} using the most recent prompt muon flux models.

\section*{Acknowledgements}
This work was supported by the National Centre for Science, Poland, grant no. 2015/18/E/ST2/00758.

%% Full authors list (ONLY FOR COLLABORATIONS)
\clearpage
\end{document}